\documentclass{aa}

\pdfoutput=1
\usepackage{pdflscape}
\usepackage{rotating}
\usepackage{longtable}
\usepackage{epstopdf}
\usepackage{subfigure}
\usepackage{multirow}
\usepackage{color}
\usepackage{colortbl}
\usepackage{ctable}
\usepackage{xspace}

\usepackage[varg]{txfonts}

\begin{document}


\title{The Supernova Remnant W44: confirmations and challenges for
cosmic-ray acceleration}

\author{M.~Cardillo\inst{1,2}
\and M.~Tavani\inst{1,2,3} \and A.~Giuliani\inst{3,4}
\and S.~Yoshiike\inst{5} \and H.~Sano\inst{5}
\and T.~Fukuda\inst{5} \and Y.~Fukui\inst{5}
\and G.~Castelletti\inst{6} \and G.~Dubner\inst{6} }

\institute{INAF/IAPS, I-00133 Roma, Italy\\
e-mail:mcardillo@roma2.infn.it
\and Dip. di Fisica, Univ. Tor Vergata, I-00133 Roma, Italy
\and CIFS-Torino, I-10133 Torino, Italy
\and INAF/IASF-Milano, I-20133 Milano, Italy
\and Department of Physics and Astrophysics, Nagoya
University, Furo-cho, Chikusa-ku, Nagoya, Aichi 464-8601, Japan
\and Instituto de Astronom\'ia y F\'isica del Espacio
(IAFE), CC.67, Suc.28, 1428, Buenos Aires, Argentina}

\date{Received / Accepted }

\abstract{
The middle-aged supernova remnant (SNR) W44 has recently attracted attention because of its relevance regarding the origin of Galactic cosmic-rays.
The gamma-ray missions AGILE and Fermi have established, for the first time for a SNR, the
spectral continuum below 200 MeV which can be attributed to neutral pion emission. Confirming the hadronic origin of the
gamma-ray emission near 100 MeV is then of the greatest importance. Our paper is focused on a global re-assessment of all
available data and models of particle acceleration in W44, with the goal of determining on a firm ground the hadronic and leptonic
contributions to the overall spectrum. We also present new gamma-ray  and CO NANTEN2 data on W44, and compare them with recently published AGILE and
Fermi data. Our analysis strengthens previous studies and observations of the W44 complex environment and provides new information  for a
more detailed modeling. In particular, we determine that the average gas density of the regions emitting 100 MeV- 10 GeV gamma-rays
is relatively high ($n \sim 250-300$ cm$^{-3}$). The hadronic interpretation of the gamma-ray spectrum of W44 is viable, and supported by strong
evidence. It implies a relatively large value for the average magnetic field ($B \geq 10^{2}$ $\mu$G) in the SNR
surroundings, sign of field amplification by shock-driven turbulence. Our new analysis establishes that
the spectral index of the proton energy distribution function is $p_{1}=2.2\pm0.1$ at low energies and $p_{2}=3.2\pm0.1$ at high energies.
We critically discuss hadronic versus leptonic-only models of emission taking into account simultaneously radio and gamma-ray data. We find that the
leptonic models are disfavored by the combination of radio and gamma-ray data. Having determined the hadronic nature of the gamma-ray emission on firm
ground, a number of theoretical challenges remains to be addressed.}


\titlerunning{The SNR W44}
\authorrunning{M.Cardillo}

\maketitle

\section{Introduction}
Cosmic-rays (CRs) are highly energetic particles (with kinetic energies up to $E=10^{20}$ eV) mainly
composed by protons and nuclei with a small percentage of electrons (1$\%$). Currently,
the CR origin is one of the most important problems of high-energy astrophysics, and the issue is the subject of 
very intense research \citep{fermi49,ginzburg64,berezinskii90}. For recent reviews see \citet{helder12}
and \citet{aharonian12}. Focusing on CRs produced in our Galaxy (energies up to the so called ``knee'', $E=10^{15}$~eV), strong shocks in Supernova Remnants
(SNRs) are considered the most probable CR sources \citep[e.g.,][]{ginzburg64}. This hypothesis is supported by several ``indirect'' signatures indicating
the presence of ultra-high energy electrons \citep[recent review in] []{vink12}. However, the final proof for the origin of CRs up to the
knee can only be obtained through two fundamental signatures. The first one is the identification of sources emitting a 
photon spectrum up to PeV energies. The second one is the detection of  a clear gamma-ray signature  of $\pi^{0}$ decay in Galactic sources. Both
indications are quite difficult to obtain. The ``Pevatron'' sources are notoriously hard to find \citep[see][for a
review]{aharonian12}, and the neutral pion decay signature is not easy to identify because of the possible contribution from
co-spatial leptonic emission. Hadronic (expected to produce the $\pi^{0}$ decay spectral signature) and leptonic components can in principle be distinguished 
in the 50-200 MeV energy band, where they are expected to show different behaviors.\\
Over the last five years AGILE and Fermi, together with ground telescopes operating in the TeV energy range (HESS, VERITAS and
MAGIC), collected a great amount of data from SNRs
\citep{abdo09_W51,abdo10_CasA,abdo10_IC443,abdo10W9b,abdo10_W44,abdo10_W28,abdo11_1713,
acciari09_IC443,tavani10_IC443,acciari10_CasA,acciari11_tycho,aharonian01_CasA,aharonian07_1713,aharonian08_W28,aleksic12_W51,giordano12_tycho,
giuliani10_W28,hewitt12_PuppisA,katsuta12_S147,lemoine12_RCW86}
providing important information and challenging theoretical models. For example, most of the observed SNRs appear to have a spectrum steeper than the
one expected from linear and non-linear diffusive shock acceleration models (DSA) of index near 2
\citep[and possibly convex spectrum][]{bell87,malkov01,blasi05}. W44 is one of the most interesting SNRs observed so
far; it is a middle-aged SNR, bright at gamma-ray energies and quite close to us. Its gamma-ray spectral index (indicative of the underlying proton/ion
distribution in the hadronic model) is $p\sim3$, in apparent contradiction with DSA  models. W44 is therefore an ideal system to
study CR acceleration in detail. The AGILE data analysis of this remnant provided for the first time information below $E=200$~MeV, showing the low-energy
steepening in agreement with the hadronic interpretation \citep{GiuCaTa11}. Recently, an analysis of Fermi-LAT data confirms these
results \citep{ackermannW44}.\\
In this paper, we present a new analysis of AGILE data together with a re-assessment of CO and radio data on W44. We also compare our results with those
obtained from Fermi-LAT data. In section~\ref{SNRW44}, we summarize the most relevant facts about W44, and in section~\ref{newAGILE}, we present an
updated view on the AGILE gamma-ray data and on the CO and radio data of this SNR.
In section~\ref{modeling}, we discuss hadronic and leptonic models in the light of our refined analysis. The implications of this work are discussed in
section~\ref{Discussion}. We provide relevant details about our modeling in the Appendices.
\begin{figure*}[!ht]
 \begin{center}
 
 \includegraphics[scale=1.8]{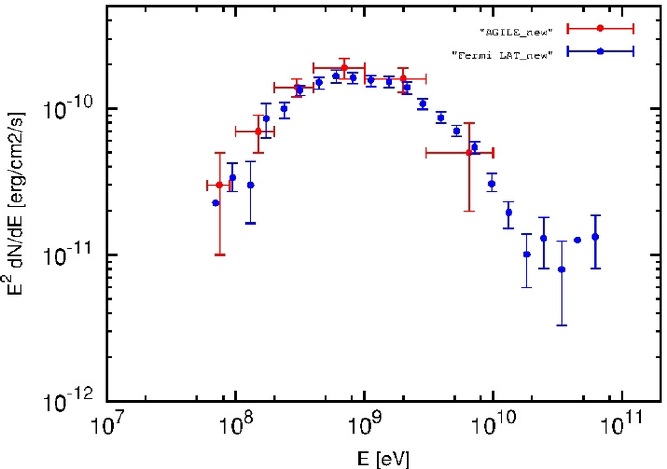}
\end{center}
\caption{AGILE new gamma-ray spectrum of SNR~W44 (red data points)
superimposed with the Fermi-LAT data from \cite{ackermannW44} (blue data points).}
\label{spectra}
\end{figure*}

\section{The supernova remnant W44}  
\label{SNRW44}

W44 is a middle-aged ($\sim$20,000 yrs old) SNR
located in the Galactic Plane ($l,b$)=~($34.7,-0.4)$, at a
distance $d\sim3.1$~kpc \citep[][\cite{fang13_W44} report 1.9 kpc.]{clark76,wolszczan91}. 
Multiwavelength observations revealed interesting features.
In the radio band, W44 shows a quasi-elliptical shell \citep[][and
references therein]{castelletti07}; the radio shell asymmetry is
probably due to expansion in an inhomogeneous ISM. In the
North-West side of the remnant, in correlation with a peak of the
radio emission, there is bright [SII] emission characteristic of shock-excited radiative 
filaments \citep{giacani97}. In the South-East side, instead,
there is a molecular cloud (MC) complex embedded in the SNR shell and interacting
with the source \citep{wootten77,rho94}. OH masers (1720~MHz)
found in correspondence with the SNR/MC region, confirm their
interaction \citep{claussen97,hoffman05}. \cite{wolszczan91}
reported the discovery of the radio pulsar PSR~B1853+01,
located in the South part of the remnant and surrounded by a
cometary-shaped pulsar wind nebula (PWN) \citep{frail96}. This system, however, does not appear to be correlated with the
detected gamma-ray emission. X-ray observations of W44  by
the Einstein Observatory \citep{watson83} showed centrally 
peaked emission, later confirmed by Chandra data
\citep{shelton04}. \\
The first report on W44 in the gamma-ray
band was by Fermi-LAT \citep{abdo10_W44} that  showed a GeV emission morphology in apparent good
correlation with the radio shell. The Fermi-LAT energy power spectrum of W44 showed a prominent peak near 1
GeV and a clear decrease at higher energies with a quite steep
spectrum of photon index near 3 \citep{abdo10_W44}. Early
processing of Fermi-LAT data had a low-energy threshold of 200
MeV, thus limiting its ability to identify a neutral pion signature. 
It is then not
surprising that in addition to hadronic models also leptonic
models predicting bremsstrahlung emission below 200 MeV could not
be excluded. The relatively large  gamma-ray brightness of W44 and the
good spectral capability of AGILE near 100 MeV
\citep{tavani08,vercellone08,vercellone09} stimulated a thorough
investigation of this supernova remnant with the AGILE data.
The AGILE gamma-ray spectrum in the range 50 MeV - 10 GeV confirms
the high-energy steep slope up to 10 GeV, and, remarkably,
identifies for the first time a spectral decrease below 200 MeV as
expected from neutral pion decay \citep[][hereafter
G11]{GiuCaTa11}.
In the analysis of G11, both leptonic and hadronic models
were considered to fit both
 AGILE and Fermi-LAT data. Proper consideration was given to
 the constraints derived from VLA radio data and
NANTEN CO data for the ambient magnetic field and density,
respectively. In G11, the best model was determined to be dominated by
hadronic emission with a proton distribution of spectral index $p_{2}=3.0 \pm 0.1$ and a low-energy cut-off at $E_{c}= 6\pm 1$~GeV.
The  W44 gamma-ray morphology determined by AGILE agrees well with the 
emission detected by Fermi-LAT below 1 GeV.
Furthermore, a correlation of gamma-ray emission with 
CO emission is observed, 
indicating that most of the gamma-ray emission can be associated with the SNR/MC interaction.
A possible large-scale influence of escaping particles
 accelerated at the W44 SNR shock was studied by
\cite{uchiyama12W44} who noticed the existence of far and bright gamma-ray bright MCs.
A new important contribution was recently produced by the
Fermi-LAT team that revisited the gamma-ray emission from W44
\citep[][hereafter A13]{ackermannW44}. This work was motivated
also by the improvement in the LAT data analysis that permits a
better study of the spectrum near 100 MeV
\citep{ackermannpass7}. The new gamma-ray spectrum of W44 by
Fermi-LAT fully confirms the AGILE spectrum below 200 MeV
\citep[][for a comparison of AGILE and new Fermi-LAT data, see
Fig. \ref{old_data} in Appendix A]{ackermannW44}. The
analysis in A13 tends to exclude a leptonic-only contribution to
the gamma-ray emission because it requires a very large density ($n\sim650 \, \rm cm^{-3}$).
Their best hadronic model, with an \textit{assumed}  
surrounding medium density $n\sim100$ cm$^{-3}$, is based on a smoothed broken power-law hadronic distribution with a
break energy $E_{br}=22$~GeV and indexes $p_{1}=2.36$ for
$E<E_{br}$, and $p_{2}=3.5$ for $E > E_{br}$. Model
parameters in A13 differ from those considered earlier in \cite{abdo10_W44}.
Apparently, in the hadronic modeling of A13 bremsstrahlung
emission is not considered to be relevant, even though in
principle this process could provide a non-negligible
contribution to the gamma-ray emissivity.
An important feature of the SNR~W44 spectrum is
its slope at GeV energies: the index $p\sim3$
is substantially steeper than the range plausibly expected in linear and non-linear DSA
 models.
In \cite{malkovW44} this spectral feature is explained
by Alfv\'en damping in the presence of a relatively
large-density medium where acceleration occurs.
The W44 environment is quite 
challenging in its morphology, and requires a reanalysis of its
properties in the context of the crucial implications for the
acceleration mechanism of CRs.
We present here a new analysis of AGILE data together with a
revised assessment of the W44 surrounding environment based on
new CO data obtained from the NANTEN2 telescope.

\begin{figure*}[!ht]
\centering
 \subfigure{\includegraphics[scale=1.3]{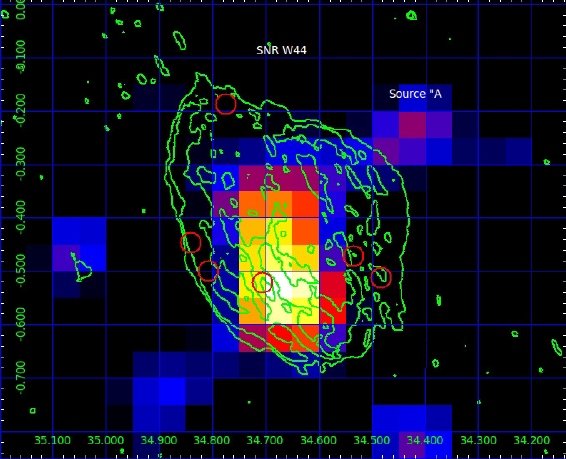} }\qquad
  \subfigure{\includegraphics[scale=3.3]{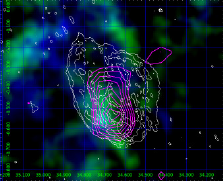} }\\
 \subfigure{\includegraphics[scale=4]{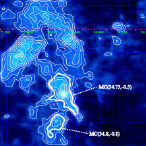}}\qquad
 \subfigure{\includegraphics[scale=2.8]{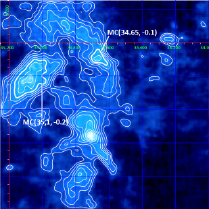}}\\
 \subfigure{\includegraphics[scale=1.1]{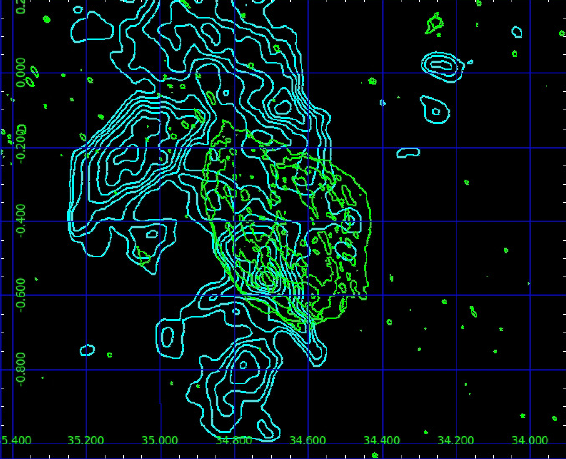}}\qquad
 \subfigure{\includegraphics[scale=1.1]{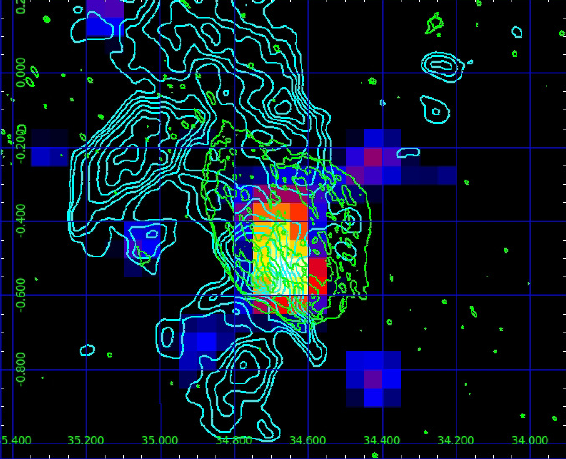}}
\caption[width=.4\textwidth]{\textbf{(Upper Left Panel)}: AGILE gamma-ray intensity map
(in Galactic coordinates) of the W44 region
in the energy range 400~MeV-10~GeV, obtained integrating all available data collected during the period
May 2007 and June 2012.
Pixel size is 0.05$^{\circ}$x 0.05$^{\circ}$ with \rm  a 3-bin Gaussian smoothing. Green contours show
the 324 MHz radio continuum flux
density detected by the Very Large Array \citep{castelletti07} and red circles indicate detected OH
masers \citep{claussen97}.
\textbf{(Upper Right Panel)}: combined CO data from the NANTEN2 observatory superimposed with the
AGILE gamma-ray data contours (magenta) above 400 MeV of
the W44 region (map in Galactic coordinates) and VLA contours (white).
CO data have been selected in the velocity range 40-43~km~s$^{-1}$, corresponding to a kinematic
distance
compatible with the W44 distance. \textbf{(Medium Left and Right Panels)}: NANTEN2 CO integrated
maps with CO contours (40-43 km/s).
Thick white lines show the four CO peaks. \textbf{(Bottom Left Panel)}: NANTEN2 CO integrated contours
(40-43 km/s, cyan) and VLA contours (green).
\textbf{(Bottom Right Panel)}: AGILE intensity map (400~MeV-10~GeV) with NANTEN2 (cyan) and VLA
(green) contours.}
\label{maps}
\end{figure*}

\section{New AGILE data analysis}
\label{newAGILE}

We performed a global reassessment of the AGILE  data on W44, including new gamma-ray data
obtained until June, 2012. The new  data were obtained  using the updated AGILE data
archive, available at the ASDC site (www.asdc.asi.it).
The analysis procedure is the same described in G11\footnote{For more details about the
background model, software and likelihood technique, see
\cite{giuliani04,giuliani06,bulgarelli12,chen13}.} except for the map
bin-size (that is now wider than before for an analysis
focused on extended features) and the substantially more extended
observing period.


\subsection{Morphology}

The upper left panel of Fig.~\ref{maps} shows the W44 AGILE
gamma-ray map in the 400-10000 MeV energy range\footnote{The
different morphology of the gamma-ray emission compared to
that presented in G11 is influenced by 
binning. In G11 we used $0.02^{\circ}$ x $0.02^{\circ}$ bins, instead here we
make the choice of a  $0.05^{\circ}$ x $0.05^{\circ}$ binning.}
with radio contours from VLA (green contour levels). The upper right panel shows the NANTEN2 telescope
CO map in two velocity channels, 41 and 43 km/s, with AGILE
(magenta) and VLA (white) contours. Gamma-ray emission appears to be mostly concentrated near a
high-density region, (bottom panel of the Fig.~\ref{maps}), indicating that most of the
W44 gamma-ray emission is coincident with  a site of 
SNR/MC interaction. This MC could be at some distance from the
remnant or embedded in it.
The CO maps (Fig.~\ref{maps}, medium panel)
show the presence of a large molecular cloud complex with four
different peaks at (34.8,-0.8), (34.75,-0.5), (35.1,-0.2), and (34.65,-0.1), respectively, 
indicated by the thick white lines.
Each of these peaks reaches densities of about $10^3 \, \rm cm^{-3}$, for an estimated average density in the SNR shell of  
$n_{av} \sim 200 \, \rm  cm^{-3}$ \citep{yoshiike13_W44}. A good
correlation with the gamma-ray emission is in correspondence with
the peak at (34.7,-0.5).


\subsection{Spectrum}

Fig.~\ref{spectra} shows the AGILE gamma-ray spectrum together with the
recently updated Fermi-LAT data from \cite{ackermannW44}.
The AGILE spectrum is composed by six energy bins between 50~MeV
and 10~GeV and our error-bars takes into account statistical errors\footnote{The systematic errors in the canonical energy band (100 MeV-3 GeV) are of order of $20\%-30\%$ of the statistical errors \citep{chen13}.}. The measured flux of the source above 400 MeV is
$F=(23\pm2)\times10^{-8}$~ph~cm$^{-2}$~s$^{-1}$.
We notice the good agreement between the two spectra.
Especially important is the confirmation of the drastic spectral
decrement below 200 MeV, a crucial feature that will be discussed
below. Both AGILE and Fermi-LAT spectra differ somewhat from the
previously published spectra in G11 and \cite{abdo10_W44}
(see Appendix~A).

\section{Modeling}
\label{modeling}

\label{hadronic_models}
We model the radio, AGILE and Fermi-LAT spectral data  by hadronic and leptonic-only scenarios, taking into account the new NANTEN2 CO data providing a value for the ISM density in the SNR surroundings\footnote{\cite{yoshiike13_W44} use a H2/CO ratio equals to $X_{CO}=1.56\times10^{20}$ $\rm cm^{-2}/K/km/s^{-1}$ (see also \cite{hunter97}) that we use as a reference value. This value of $X_{CO}$ is known to be uncertain within a factor of at least 2 (see, e.g., \cite{strong04}).}, $n_{av}\simeq 250\,\rm cm^{-3}$ \citep{yoshiike13_W44}. This value of
the average gaseous density surrounding the gamma-ray emission is substantially larger than the one assumed in G11 and A13
($n=100 \, \rm cm^{-3}$). Since the AGILE gamma-ray emission is strongly correlated with one of the CO peaks, in the following we consider an average density
$n\simeq 300\pm 50$ cm$^{-3}$ $>n_{av}$. In modeling the spectra, we consider the most statistically significant Fermi-LAT data up to 50 GeV.

\subsection{Hadronic Models}
  \begin{figure}[!h]
   \begin{center}
  \includegraphics[scale=1.2]{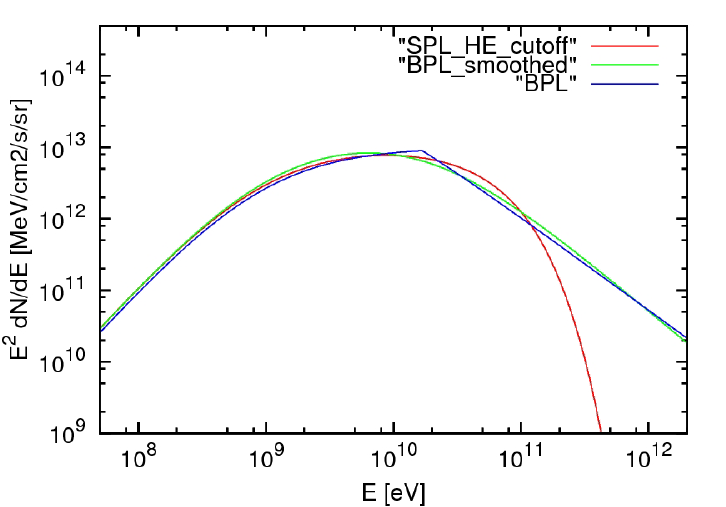}
   \caption{Particle total energy distributions for our best hadronic models plotted vs the kinetic energy: a simple power-law with a high-energy cut-off at
$E_{c}=45$ GeV (Eq.~\ref{simple_HE}, red), a smoothed broken power-law with $E_{br}=16$ GeV (Eq.~\ref{broken}, green) and
broken power-law  with $E_{br}=20$ GeV (Eq.~\ref{broken_2}, blue).}. \label{distributions}
   \label{fig_hadronic_distributions}
   \end{center}
  \end{figure}

\begin{table*}[!th]
\caption{Hadronic model parameters: p$_{1}$ is the proton spectral index before the break, p$_{2}$ is the proton spectral index above the break, $E^{p}_{br}$ is the proton break energy, $E^{p}_{c}$ is the proton cut-off energy, $K_{p}$ and $K_{e}$ are proton and electron normalization constants, and $\frac{\chi^{2}}{n-1}$ is the reduced chi-square.}
\footnotesize
\centering
\label{table_hadronic}
\begin{tabular}{l c c c c c c c}
\hline\hline
\textbf{Models}              &\textbf{$p_{1}$}        & \textbf{$p_{2}$}        & $\mathbf{E^{p}_{br}}$  & $\mathbf{E^{p}_{c}}$ & $\mathbf{K_{p}}$                         &  $\mathbf{K_{e}}$               & $\mathbf{\chi^{2}/n-1}$         \\
                       		   &                                   &                                      &   \textbf{[GeV]}               & \textbf{[GeV]}               &  \textbf{[1/MeV/cm$^3$]}                 &  \textbf{[1/MeV/cm$^3$]} &                \\
\hline                             
                             	   &                        &                         &                        &                      &                                          &                          &                  \\
\textbf{H1}                  	  &       $2.0\pm0.1$      &       -                 &     -                  &   $45\pm1$           &     $2\times10^{-14}$                   &   $4\times10^{-14}$       &       2           \\
                           		  &                        &                         &                        &                      &                                          &                          &                   \\  
                         		    &                        &                         &                        &                      &                                          &                          &                 \\
\textbf{H2}                 	&      $1.7\pm0.1$       &  $3.5\pm0.1$            &     $16\pm1$           &       -              &  $3.9\times10^{-14}$                     &   $7\times10^{-14}$       &           1.8       \\
                           		  &                        &                         &                        &                      &                                          &                          &                 \\
                          		   &                        &                         &                        &                      &                                          &                          &            \\
\textbf{H3}                  &     $2.2\pm0.1$        &  $3.2 \pm0.1$           &      $20\pm1$          &   -                  &$K_{1}\sim1.8\times10^{-13}$, $K_{2}\sim1.5\times10^{-13}$  & $1.5\times10^{-12}$         &       1.5            \\
                          		   &                        &                         &                        &                      &                                          &                          &               \\
\hline
 \end{tabular}
\end{table*}

\begin{figure*}[!ht]
\centering
 \subfigure{\includegraphics[scale=1.1]{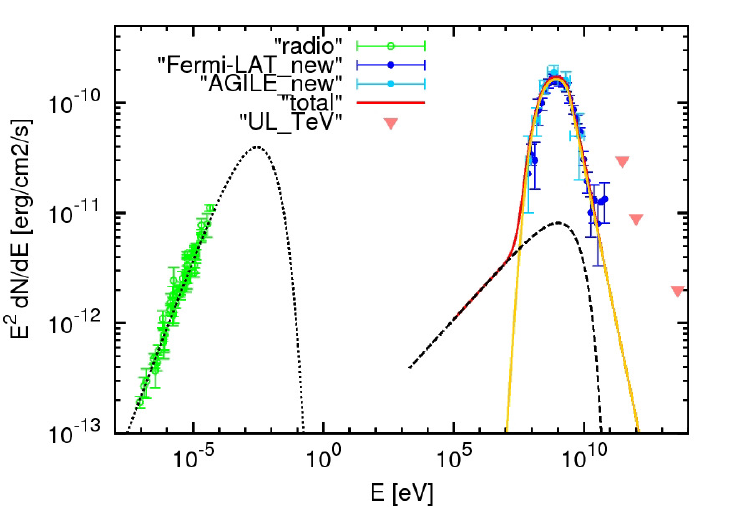}}\qquad
 \subfigure{\includegraphics[scale=1.1]{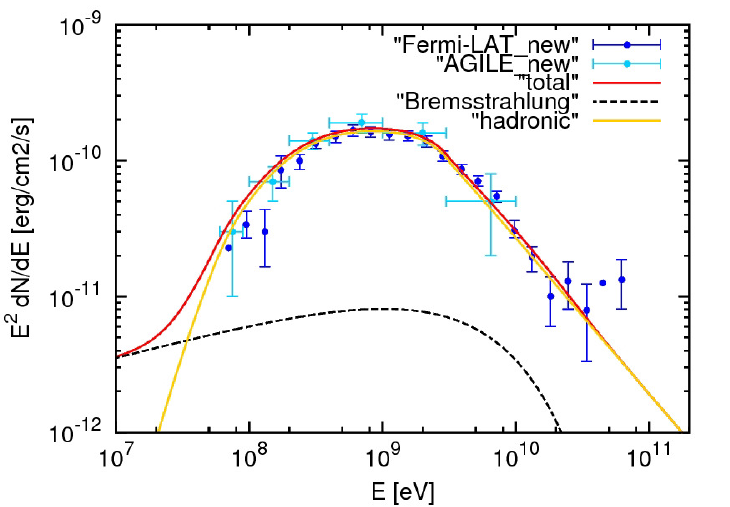}}
 \caption{Our best hadronic model, H3, of the broad-band spectrum of the SNR~W44 superimposed with radio \rm (data points in green color)
 and gamma-ray data of Fig.~\ref{spectra} (in blue and cyan color).
Proton distribution in Eq.~\ref{broken_2} with index
$p_{1}=2.2\pm0.1$ (for $E<E_{br}$), $p_{2}=3.2\pm0.1$ (for
$E>E_{br}$) where \textit{$E^{p}_{br}$}= 20 GeV. This model is
characterized by \textit{B} =210~$\mu$G and \textit{n}
=300~cm$^{-3}$. The yellow curve shows the neutral pion emission
from the accelerated proton distribution discussed in the text.
The black curves show the electron contribution by synchrotron
(dot) and bremsstrahlung (dashed) emissions; the IC contribution is
negligible. The red curve shows the total gamma-ray emission from
pion-decay and bremsstrahlung. \textbf{(Left Panel)}: SED from
radio to gamma-ray band. \textbf{(Right Panel)}: only gamma-ray
part of the spectrum.}
\label{hadronic}
\end{figure*}
We assume that the gamma-ray emission spectrum is due to the
combined contribution of hadronic $\pi^{0}$ emission and leptonic
bremsstrahlung emission, considering the proton component as the
main one. For hadronic emission, we use the formalism explained in
\cite{kelner06} that is a good approximation of the exact
solution. We consider a proton distribution in total energy $E$,
rather than in kinetic energy $E_{k}=E-m_{p}c^{2}$, following \cite{simpson83} and
\cite{dermer86} but with $\delta$-function approximation for the cross section \citep{aharonian04}.
This approximation provides that a fixed
fraction of proton energy is converted to $\pi^{0}$ energy. Even if the distribution is broad, this method gives results with good accuracy as long as the proton spectrum is smooth and broad (e.g., power-law).
We fit the gamma-ray data assuming different
types of proton distributions (see Fig. \ref{distributions}):
\begin{itemize}
 \item a simple power-law with a high-energy cut-off (\textbf{model H1}):
\begin{equation}
 \frac{dN_{p,1}}{dE}= K_p \left( \frac{E}{E^p_c} \right)^{-p_{1}} e^{-\frac{E}{E^p_c}}
 \label{simple_HE}
\end{equation}
\item a smoothed broken power-law (\textbf{model H2}):
\begin{equation}
 \frac{dN_{p,2}}{dE}= K_p \left(   \frac{E}{E^p_{br}}    \right)^{-p_1} \left(    \frac{1}{2} \left( 1 + \frac{E}{E^p_{br}}  \right)  \right)^{p_1 - p_2}
 \label{broken}
\end{equation}
\item a broken power-law (\textbf{model H3}):
\begin{equation}
\frac{dN_{p,3}}{dE}=\left\{ \begin{array}{ll}
K_{p,1}  \left( \frac{E}{E^p_br} \right)^{-p_{1}} & \textrm{if $E<E_{br}$}\\
K_{p,2}  \left( \frac{E}{E^p_br} \right)^{-p_{2}} & \textrm{if $E>E_{br}$}
\end{array} \right.
 \label{broken_2}
\end{equation}
\end{itemize}

For leptons, in all hadronic models we used a simple power-law with a high energy cut-off:
\begin{equation}
 \frac{dN_{e}}{dE}= K_e  \left( \frac{E}{E^e_c} \right)^{-p'} e^{-\frac{E}{E^e_c}}
 \label{simple_el}
\end{equation}
We fix only the parameters for which we have solid observational evidence; the average medium density, $n=300 \, \rm  cm^{-3}$ and the radio
spectral index, $p'=1.74$.
We vary all other parameters such as
 the normalization constants, $K_{p}$ and $K_{e}$, and the
cut-off and break energies, $E_{c}$ and $E_{br}$. Our results are
summarized in Table \ref{table_hadronic} where we show the best models
obtained according to the standard chi-square minimization test \citep{Taylor2000}. Every model
is discussed individually in Appendix \ref{appendixB}.
Here we present the properties of our best hadronic model H3. This is
characterized by the distribution in Eq. \ref{broken_2} with $p_{1}=2.2\pm0.1$ (for
$E<E_{br}$), $p_{2}=3.2\pm0.1$ (for $E>E_{br}$), and an energy break $E^{p}_{br}=20$~GeV. The leptonic contribution to this
model is given by a simple power-law for the electrons, with $p'=1.74$, and $E^{e}_{c}=12$~GeV (see Fig.~\ref{hadronic}). This
model provides a proton energy $W^{p}=5\times10^{49}$ erg and requires an average magnetic field in the emission region, $B=210$ $\mu$G.\\
In our calculations we do not consider the so called "nuclear enhancement factor" \citep{dermer12} that takes into account helium contribution to the gamma-ray spectrum and this is of the order of 2. However, the only change
due to this factor is a reduced proton energy density; our most important results and conlusions about spectral indices and parameter estimation are not affected in any way.


\subsection{Leptonic-only Models}
\label{leptonic-only models}
\begin{figure*}[!ht]
 \centering
  \subfigure{\includegraphics[scale=1.2]{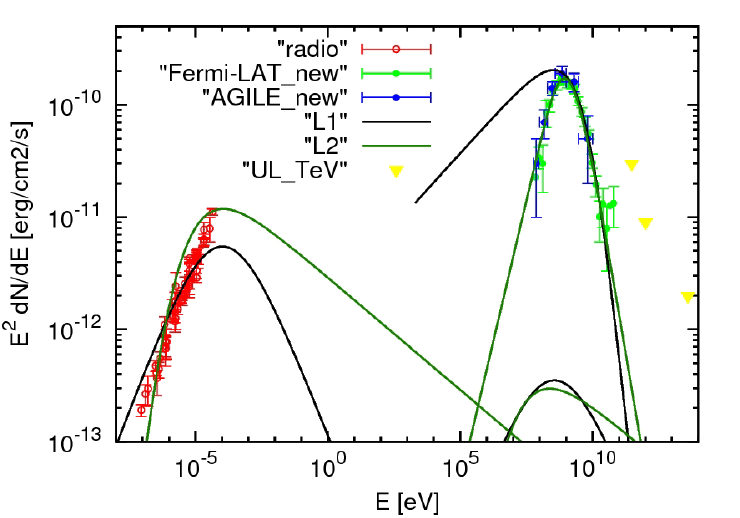}}\\
  \subfigure{\includegraphics[scale=1.15]{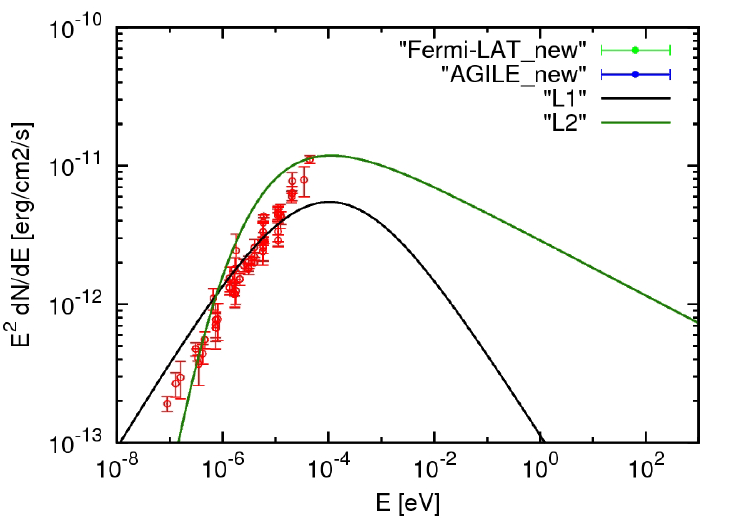}}\qquad
  \subfigure{\includegraphics[scale=1.15]{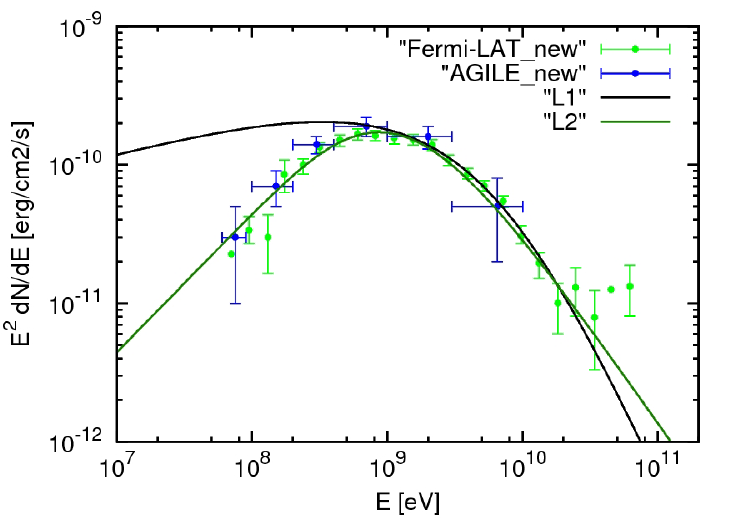}}\\
   \caption[width=.4\textwidth]{Photon spectra obtained from the two leptonic-only models, L1 (green) and L2 (black), based on a broken power-law electron
   distribution. The density is fixed at $n=300$ cm$^{-3}$. In the model L1 the electron index is fixed at $p'_{1}=1.74$, obtained from radio data, and it
   provided $p'_{2}=4.2$, $B=25$ $\mu$G and $E^{e}_{br}=8$ GeV. In the model L2, instead, we obtained $p'_{1}=-2.5$, $p'_{2}=3.4$, $B=40$ $\mu$G
   and $E^{e}_{br}=500$ MeV.}
\label{leptonic}
 \end{figure*}

 It is important to test the viability of leptonic-only models of gamma-ray emission from W44.
 We use the general expression for electron radiative processes as in \cite{blumenthal70}.
 \begin{itemize}
  \item Synchrotron emission - For a power-law electron distribution, it is convenient to use a $\delta$-function approximation \citep[e.g.,][]{longair11}
  that can be expressed as:
   \begin{equation}
    \frac{dN}{dE_{\gamma}}=4\pi \frac{1}{E_{ph}}\sigma_{th}U_{B}F_{e}(E_{e})dE_{e}\,\,[\frac{1}{MeV\,s\,cm^{-3}}]
   \end{equation}
   where $U_{B}=\frac{B^{2}}{8\pi}$ MeVcm$^{-3}$ is magnetic energy density, $\sigma_{th}$ is Thompson cross section,
   $dE_{e}=\frac{m_{e}c^{2}}{2}\left(\frac{3}{4}\frac{1}{E_{\gamma}E_{ph}}\right)$ and $E_{ph}=\frac{Be}{2\pi m_{e} c}h$ is the initial photon energy.
   For a power-law electron distribution proportional to $E^{-p'}$, the photon energy distribution is then proportional to
   $E^{-(\frac{p'-1}{2})}$.
  \item Bremsstrahlung emission - We used the general expression from \cite{blumenthal70}:   
   \begin{equation}\begin{split}
    &\frac{dN}{dE}=\alpha r_{0}^{2}E^{-1}n\int dE_{e}F_{e}(E_{e})E_{e}^{-2}\\
    &\left[\left(2E_{e}^{2}-2E_{e}E+E^{2}\right)\phi_{1}-\frac{2}{3}E_{e}\left(E_{e}-E\right)\phi_{2}\right]\,\,[\frac{1}{MeV\,s\,cm^{-3}}]
   \end{split}\end{equation}
   where $n$ is the density, $\phi_{1}$ and $\phi_{2}$ are functions of electron energies, $\alpha$ is the fine structure constant and $r_{e}$ is the electron
   classical radius. This can be used both in a totally ionized medium (weak shielding) and in the presence of neutrals (strong shielding);
   difference is only a logarithmic factor. For a power-law electron distribution proportional to $E^{-p}$, the photon energy distribution is then
   proportional to $E^{-p'}$.
 \item Inverse Compton emission
  \begin{equation}
    \frac{dN}{dE}=4\pi \frac{1}{E_{ph}}\sigma_{kn}U_{ph}F_{e}(E_{e})dE_{e}\,\,[\frac{1}{MeV\,s\,cm^{-3}}]
   \end{equation}
   where $U_{ph}$ energy density of the radiation field, $\sigma_{kn}$ is Klein-Nishina cross section,
   $dE_{e}=\frac{m_{e}c^{2}}{2}\left(\frac{3}{4}\frac{1}{E_{\gamma}E_{ph}}\right)$ and $E_{ph}$ is the interstellar radiation field initial photon energy.
 \end{itemize}
 In order to fit both gamma-ray and radio data, we consider a smoothed broken power-law leptonic distribution:
 \begin{equation}
  \frac{dN_{e,2}}{dE}= K_p \, \left( \frac{E}{E^e_{br}} \right)^{-p'_1} \left( \frac{1}{2}
  \left( 1 + \frac{E}{E^e_{br}}  \right)  \right)^{p'_1 - p'_2}
 \end{equation}
 Our assumption is that the same electron population produces both the gamma-ray and the radio fluxes through bremsstrahlung and synchrotron emissions,
 respectively. The spatial co-existence of radio filaments and sites of gamma-ray emission justifies this hypothesis.
 We fix the gaseous density value, $n=300 \, \rm cm^{-3}$, from NANTEN2 data.\\
 Our first leptonic-only model was developed (\textbf{L1}, see Table~\ref{table_leptonic}) fixing the electron spectral index at the value
 found from radio data analysis by \citet{castelletti07}, $p'_{1}=1.74$ for $E<E_{peak}$. We found a high energy electron spectral index $p'_{2}=4.2\pm0.1$
 above an energy break $E^{e}_{br}\sim 8$ GeV and a magnetic field $B\sim25$ $\mu$G. However, fixing $p'_{1}=1.74$, we can fit radio synchrotron data but we
 cannot fit in any way the low-energy gamma-ray data (see Fig.~\ref{leptonic}).\\
 The second leptonic-only model was developed in order to fit gamma-ray data with the Bremsstrahlung emission (\textbf{L2},
 see Table~\ref{table_leptonic}), changing the electron spectral index. We can fit low-energy gamma-ray data with an index $p'_{1}=-2.5\pm0.1$ for
 $E<E^{e}_{br}$, very hard to explain. The other parameters found are $p'_{2}=3.4\pm0.1$ for $E>E^{e}_{br}$, $E^{e}_{br}\sim 500$ MeV and $B\sim40$ $\mu$G.

\begin{table*}[!ht]
\caption{Leptonic model parameters: p'$_{1}$ is the electron spectral index before the break, p'$_{2}$ is the electron spectral index above the break, $E^{e}_{br}$ is the electron break energy, and $K_{e}$ is the electron normalization constant}.
\footnotesize
\centering
\label{table_leptonic}
\begin{tabular}{l c c c c}
\hline\hline
\textbf{Models}              &\textbf{$p'_{1}$}       & \textbf{$p'_{2}$}      & $\mathbf{E^{e}_{br}}$  &  $\mathbf{K_{e}}$           \\
                             &                        &                         &   \textbf{[GeV]}       &  \textbf{[1/MeV/cm$^{-3}$]}           \\
\hline                            
                             &                         &                         &                       &            \\
\textbf{L1}                  &      1.74              &       $4.2\pm0.1$       &     $8\pm1$            &     $4\times10^{-14}$       \\
                             &                        &                         &                        &      \\  
                             &                        &                         &                        &           \\
\textbf{L2}                  &      $-2.5\pm0.1$      &        $3.4\pm0.1$      &       $0.5\pm0.1$      &      $1\times10^{-11}$                \\
                             &                        &                         &                        &                \\ 
\hline
 \end{tabular}
\end{table*}

\section{Discussion}
\label{Discussion}


\subsection{Models}
\label{models}

Gamma-ray emission from SNRs can be produced in general by three different mechanisms: (1) relativistic bremsstrahlung
from electrons interacting with surrounding medium, (2) inverse Compton emission from electrons scattering soft
photons (e.g. cosmic background radiation and interstellar radiation field), and (3) proton-proton interaction
producing $\pi^{0}$ which subsequently decay into two gamma-ray photons. In order to find an unambiguous
signature of accelerated hadrons in W44, we need to clearly identify these different contributions in the high energy spectrum. As in G11, we 
 model the gamma-ray data considering all possible emission mechanisms. 
We fix two important parameters obtained from radio and mm-CO data. Multifrequency radio data \citep{castelletti07}
provide the radio photon index $\alpha=0.37$ that implies an electron index $p'=1.74$, for energies less than the synchrotron peak. By using NANTEN2
telescope data, we can also fix the SNR average density in the region of gamma-ray emission at $n_{av}= 300 \, \rm  cm^{-3}$.

\subsubsection{Leptonic-only model failure}

Our aim is to test whether a leptonic-only model can explain the gamma-ray emission from W44. We assume that the same
electron population produces both the radio and the gamma-ray emissions. We assume a broken power-law electron distribution with Inverse Compton and
Bremsstrahlung components.\\
 \begin{itemize}
  \item L1 model - We use as a parameter the index $p'_{1}=1.74$, obtained from radio synchrotron data \citep{castelletti07}. Relativistic bremsstrahlung
  has the same electron index \citep{blumenthal70}. Consequently, an index $p'_{1}=1.74$ cannot explain the low-energy gamma-ray data in any way
  (Fig.~\ref{leptonic}, black curve). Moreover, the relation between density and magnetic field (see Appendix~C) constrains the synchrotron peak; fixing the
  medium density to the average value found in \cite{yoshiike13_W44}, $n=300$ cm$^{-3}$, we cannot fit in a good way the W44 radio emission for any magnetic
  field value.
  The best model gives a $B=25$ $\mu$G. Changing the density value does not improve the fit.
  \item L2 model - In this case, we do not apply the radio constraint on the electron spectral index in order to fit the low-energy gamma-ray data.
  We find that only an index $p'_{1}=-2.5\pm0.1$ can explain the gamma-ray spectrum decay at $E<E^{e}_{br}$ with $E^{e}_{br}=500$ MeV, together an index
  $p'_{2}=3.4\pm0.1$ for $E>E^{e}_{br}$. In this case, the W44 gamma-ray emission can be explained but the
  radio synchrotron data (see Fig.~\ref{leptonic}, green curve) are in contradiction with the model. 
 \end{itemize}
Fig.~\ref{leptonic} shows the failure of these
two models\footnote{A substantial inverse Compton contribution can be easily excluded. If the soft photon source is the Cosmic Background Radiation (CBR),
we expect a second peak in the gamma-ray spectrum ($E_{peak}\sim 1$ TeV), contradicting upper-limits from Cherenkov telescopes\citep{abdo10_W44}. If soft
photons come from  the InterStellar Radiation Field (ISRF), we cannot fit synchrotron radio data in any way, using reasonable magnetic field values.},
L1 and L2, to simultaneously explain the radio and gamma-ray emission of W44.

\subsubsection{The best hadronic model: a broken power-law distribution}
We considered different possible hadronic models (Table~\ref{table_hadronic} and
Fig.~\ref{distributions}). The best one is the so called H3, a broken power-law proton distribution discussed in the following.
Features and implications of the other models, H1 and H2, are discussed in Appendix A.
In order to consider separately the low and high energy parts of the spectrum, we consider a non-smoothed broken power-law
distribution (Eq. \ref{broken_2}). In this way we can study which kind of processes affect one or the other part of the spectrum. We obtain a good model with
$p_{1}=2.2\pm0.1$ for $E<E_{br}$ and $p_{2}=3.2\pm0.1$ for $E>E_{br}$, where $E_{br}=20$ GeV. The magnetic field is of the same order as for other hadronic
models, $B=210$ $\mu$G, and the electron/proton number ratio is $\Re=0.01$ at $e\sim10$ GeV. The index found for the low-energy part is in
agreement with the one found by A13. On the other hand, the high energy index of the H3 model is substantially harder than in A13 and steeper
than the one found in our previous paper, $p_{2}=3.0\pm0.1$ \citep{GiuCaTa11}. Our analysis confirms that the SNR~W44 has a
gamma-ray spectral index near 10 GeV steeper than all other middle-aged SNRs. Interestingly, the low-energy index near 2.3 is close to the
value found in several other young SNRs \citep{abdo10_CasA,acciari10_CasA,acciari11_tycho,aharonian01_CasA,giordano12_tycho,
hewitt12_PuppisA}. This fact  can have a profound reason, and it may be related with a universal or quasi-universal injection of energetic particles by a
SNR shock.


\subsection{Hadronic models: proton energies and magnetic fields.}

An important physical consequence of hadronic models is the value of the total energy going into accelerated protons.
Considering the total particle energy, \cite{castelletti07} provided a minimum value for the total CR energy and for the
magnetic field in W44, estimated from the radio data, assuming particle and magnetic energy equipartition;
$U_{min}=5.8\times10^{49}\,erg$ and $ B_{min}=13\,\mu G$. For the equipartition assumption, $U_{min}=2U_{B}=2U_{CR}$.
Consequently, far away from equipartition, magnetic energy should be greater than particle
energy or viceversa \citep{longair11}. We can use the equipartition values for magnetic and particle energies found by \cite{castelletti07}, to
expose a contradiction in their relation. Our models, except for the simple power-law with a high energy cut-off, provide a
total energy in accelerated protons lower than the one calculated with the equipartition assumption, implying magnetic fields with higher values.
It is essential to obtain a good estimate of the magnetic field in regions of interest in W44.
In our models, in order to fit the gamma-ray data, we explicitly consider also the bremsstrahlung contribution by electrons. 
This approach provides a constraint on the magnetic field. Assuming that synchrotron and bremsstrahlung emissions are originated by the same electron
population, both processes depend on the electron density; the higher its value, the higher their
emissivity. Consequently, in order to obtain a small bremsstrahlung contribution to the gamma-ray emission with a fixed target
density, the magnetic field has to be enhanced in order to obtain the correct synchrotron emission, and viceversa. The final result
is that, regardless of the hadronic model, we can fit the radio data only considering a magnetic field
$B\sim10^{2}~$ $\mu$G $>>B_{min}$ (see Table \ref{table}). This fact implies that the magnetic energy should be the main contribution to the total energy.
Consequently, our model H1 can be excluded because both magnetic and particle energies are grater than the equipartition values.
Large values of the magnetic field ($B\sim0.2$ mG) in W44 were deduced by \cite{claussen97} in regions near the
detected OH masers. Interestingly, from Fig.~\ref{maps}, we find that the gamma-ray emission detected by AGILE overlaps with one of the OH maser regions.
In order to obtain a large local magnetic field (i.e., substantially larger than the equipartition one) an efficient amplification mechanism must be operating.
A possible mechanism was discussed by \cite{schure12}: a linear magnetic instability can provide the condition $\delta
\textbf{B}\sim \textbf{B}_{0}$. However, it is required that the instability continues to grow in order to explain magnetic fields greater than $100$~$\mu
G$. Identifying the physical mechanisms for magnetic field amplification can be challenging. It is also important to explain the relation between magnetic
field and density structures \citep[{e.g.,} Fig.2 in][]{schure12}. In young  SNRs where the ISM density is low ($n\sim 1 \,
cm^{-3}$), high magnetic fields are usually correlated with optical and radio filaments. In middle-aged SNRs such as W44 surrounded by high ISM densities
($n\sim10^{2} \, \rm cm^{-3}$), the magnetic field is relatively large on wider scales.

\subsection{Spectral index}
 Our best hadronic model is obtained from a non-smoothed broken power-law distribution. At
low-energies, we can fit the gamma-ray  data with a proton distribution index $p_{1}=2.2$. This value apparently agrees with
the behavior seen in younger SNRs. The difference is that in young SNRs this spectral index applies also at higher energies 
not being affected by propagation and damping mechanisms.
On the contrary, in  the middle-aged SNR~W44, at higher energies we find a proton index $p_{2}=3.2$, that is substantially steeper that the value expected
from theoretical models without damping. \cite{malkovW44} explained the W44 steep spectral index with the mechanism of Alfv\`en
damping, providing a steepening of exactly one unit. However, Alfv\`en damping, if it occurs in W44, cannot be acting all
across the proton spectrum because the deduced low-energy index
seems unaffected by it.\\ 
It is interesting to compare W44 gamma-ray spectrum with the CR particle interstellar spectrum. The interstellar cosmic-ray proton spectrum results
to be, in the momentum space, $p_{1,IS}=2.5$ below $E=6.5$ GeV and $p_{2,IS}=2.8$ above $E=6.5$ GeV \citep{dermer13};
the interstellar cosmic-ray electron spectrum, instead, seems to have an index $p'_{1,IS}=1.3$-$1.6$ below a few GeV and $p'_{2,IS}=2.1$-$2.3$ above
GeV energies \citep[][and therein]{strong11}.
Considering our best hadronic model, H3, it provides a proton index at the lowest energies, $p_{1}=2.2$, in the energy space. This implies a proton
index in the momentum space similar to the interstellar CR spectrum one, $p_{1,m}=2.4-2.5$, and the high energy proton spectral index, $p_{2}=3.2$,
results to be steeper than the interstellar CR spectrum one.
For the electrons, instead, radio data provided an index $p'=1.74$ \citep{castelletti07}, steeper than $p'_{1,IS}$ but harder than $p'_{2,IS}$.
Consequently, the CR spectral behavior in SNR~W44 is different from the one of the interstellar CR spectrum. This challenging issue requires a deeper
analysis beyond the scope of this paper. We note, however, that also in the interstellar case proton and electron spectra have different indices as well
as in W44.\\
Our spectral indices are compatible with the values found in \cite{ackermannW44}, $p_{1}=2.36\pm0.05$ for $p<p_{br}=22$ GeVc$^{-1}$, and $p_{2}=3.5\pm0.3$ for $p>p_{br}$, where a smoothed broken power-law is used and there is any consideration about the electron contributions. \cite{fang13_W44}, instead, assumes that the W44 spectral behavior is explained by diffusive shock acceleration with ion-neutral damping and consider a lower distance of the remnant ($d\sim1.9$) and a lower density ($n\sim10^{2}$ cm$^{-3}$). In this way they found avery low magnetic field ($B\sim10$ $\mu$G) and a very steep spectral index, $p_{1}\sim4.1$. A direct comparison with our results is not so trivial because of the different approaches: we, as like as \cite{ackermannW44}, begin from fitting our data and then we look for a physical explanation. \cite{fang13_W44}, instead, begins from the issue that a linear DSA theory can explain Fermi-LAT data and then obtain the parameters that are inconsistent with both our and \cite{ackermannW44} values.\\
It is interesting to remark here that data collected from the young SNRs, Tycho and Cas~A \citep{abdo10_CasA,giordano12_tycho},
show a spectral index in the range $p_{1}\approx 2.2$-$2.4$, i.e., steeper than the value $p_{1}=2$ predicted by idealized theoretical
models \citep[assumed by][]{malkovW44}. Other non-linear mechanisms modifying standard DSA, such as neutral leakage \citep[][and therein]{blasi12b} or
re-acceleration \citep[][and therein]{blasi12a}, may substantially affect SNRs in the whole range of energies.


\section{Conclusions}
\label{Conclusions}
W44 is a crucial SNR providing important information on the CR origin in our Galaxy.
However, several characteristics of this SNR  deduced by a multifrequency approach (gamma-ray spectral indexes, large
magnetic field) are challenging. 
As discussed in this paper, W44 is a relatively close and quite  bright gamma-ray source.
Therefore, an excellent characterization of its gamma-ray spectrum in the range 50-200 MeV has been possible because of the
good statistics achieved by AGILE and Fermi-LAT. In this paper we re-analyzed the spectral properties and the likelihood of
interpreting the decrement below 200 MeV as a ``pion bump''. 
We performed a re-analysis of the AGILE data, together with revisiting radio and CO data of W44.
We showed the unlikeliness of  leptonic-only models in their most natural form: electron distributions constrained by radio data cannot fit the broad-band
W44 spectrum. On the other hand, we find that both gamma-ray and radio data can be successfully modeled
by different kinds of hadronic models (H1, H2, H3).\\
Our results regarding the spectral properties of the accelerated proton/ion population by the W44 shock are in qualitative agreement with the results
of \citep{GiuCaTa11}.
We provided in this paper a broader discussion of alternatives, and specified the role played by leptons alone and jointly with protons.
In what follows, we summarize the most important physical characteristics of this source.
 \begin{itemize}
  \item \textbf{Neutral pion signature -} W44 is the first SNR clearly showing the so called
  ``pion bump'' that we expect at $E\geq67$ MeV from $\pi^{0}$-decay photons.
 The low-energy gamma-ray spectral index in our best model is $p_{1}=2.2\pm0.1$.
 This value is similar to those found in young SNRs,
indicating that the proton injection spectrum is affected by non-standard mechanisms of acceleration.  
  \item \textbf{High density of the surrounding environment -}
We determined that the average density in the W44 shell is 
 $n_{av} \sim 300 \, \rm cm^{-3}$, with $n\geq 10^{3}$~cm$^{-3}$ in
 correspondence with CO peaks (see medium panels in
 Fig.~\ref{maps}). This feature was also found in other middle-aged
 SNRs, like W51c and IC443 \citep{koo10,castelletti11} and
 explains the high gamma-ray flux detected from these sources. In
 the SNR~W28, the average density is lower, $n_{av}\approx
 5$~cm$^{-3}$ \citep{gabici09}, but gamma-ray emission was detected in good
  correlation with the two MC complexes where
 $n\approx 10^{3}$~cm$^{-3}$ \citep{giuliani10_W28}.
  \item \textbf{High magnetic field -} In W44 our best hadronic models imply a
  magnetic field  $B\geq 100$ $\mu$G, which is
 lower than the post-shock magnetic field estimated by \cite{claussen97}  from Zeeman splitting in the OH masers,
  and substantially higher than the equipartition magnetic field
  \citep{castelletti07}. However, in most of SNRs, 
  magnetic field estimations give values $B\sim10-10^{2}$ $\mu$G that are
  much higher than the average diffuse galactic value
 [see, for example, \cite{morlino12} for Tycho, and \cite{koo10}
 and \cite{tavani10_IC443} for W51c and IC443, respectively]. This
 is hardly surprising since magnetic field compression due to the shock interaction with the
 ISM leads to its amplification. 
 We need then to consider a non-linear scenario
 with a back-reaction of the accelerated particle at
 the shock \citep{bell01}. The large value for the magnetic field in W44 may be linked to the environment density value, $n_{av}\sim 300 \, \rm cm^{-3}$
  given by NANTEN2. We notice that for a lower density value, we can enhance the electron density
  and make plausible a lower magnetic field. 
  \item\textbf{ Steepness of the high energy index -} As in \cite{abdo10_W44}, G11, and A13,
W44 shows for energies above 1~GeV, a spectral index $p_{2}\sim3$ that is steeper than
the values found in other middle-aged SNRs. Alfv\`en damping in a dense environment \citep{malkovW44} is a
 mechanism for explaining this behavior, but other possibilities exist  \citep[e.g.,][]{blasi12a,blasi12b}. This is a point
 requiring deeper investigations in the future.
\end{itemize}
Our final conclusion is that W44 stands out as a crucial SNR whose gamma-ray emission can be firmly demonstrated to be of hadronic
origin. A complete understanding of the W44 features requires modeling physical processes beyond DSA. Future investigations
will have to address these issues as well as understanding W44 within the context of other SNRs.

 \begin{acknowledgements}
 We thank an anonymous referee for his/her comments that led to substantial improvements of our paper.
 We are pleased to thank F. Aharonian for extensive discussions
 that stimulated parts of the work presented in this paper.
  Research partially
 supported by the ASI grants n. I/042/10/0 and I/028/12/0 and Argentina ANPCyT and CONICET grants: PICT~0902/07, 0795/08, 0571/11, PIP~2166/08, 0736/12. 
 \end{acknowledgements}

\newpage

\section*{Appendix A - Old and new AGILE and Fermi-LAT data on W44}
\label{appendixA}

\begin{figure}[ht!]
 \centering
 \subfigure{\includegraphics[scale=1.2]{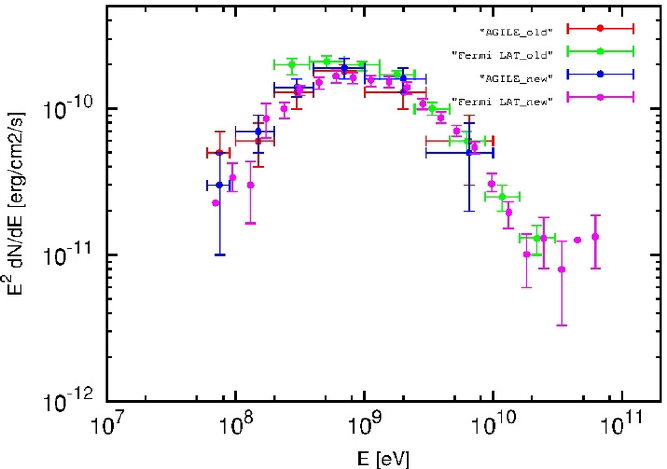}}\qquad
 \subfigure{\includegraphics[scale=1.2]{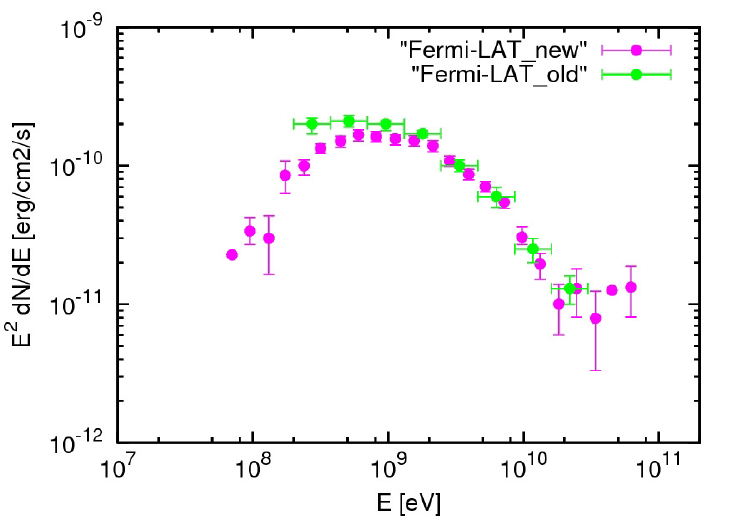}}
\caption{(\textbf{Top Panel}): AGILE (red) and Fermi-LAT (green) old spectral energy distribution (SED) of W44 \citep{GiuCaTa11,abdo10_W44}, together with
the new Fermi-LAT (magenta) \citep{ackermannW44}.The new AGILE data are shown in blue 
in Fig.~\ref{spectra}. (\textbf{Bottom Panel}): Fermi-LAT new (magenta) \citep{ackermannW44} and old (green) \citep{abdo10_W44}
spectral energy distributions. We note that there is a substantial
difference at low-energies between the two data set.} \label{old_data}
\end{figure}

Fig.~\ref{old_data} shows the old and new AGILE and
Fermi-LAT spectral gamma-ray data W44. Low-energy spectral points have been added to the Fermi-LAT
spectrum because to its recently improved analysis \citep{ackermannpass7}.
We notice that the low-energy Fermi-LAT spectrum  below 200 MeV has 
changed with respect to the previous Fermi-LAT data of
\cite{abdo10_W44}.
On the other hand, the AGILE data re-analyzed in this paper
are no so different from those previously  presented in G11
except for the lowest energy point between $50$-$100$ MeV.
This lowest-energy spectral data is lower than the one found in our previous paper, and agrees
with the Fermi-LAT spectral trend. Constraining the gamma-ray spectrum near 50 MeV is of the
greatest importance for its relevance related with a possible bremsstrahlung contribution.

\section*{Appendix B - Other hadronic models}
\label{appendixB}

\subsection*{Simple power-law with a high-energy cut-off}

\begin{figure}[!h]
 \begin{center}
 \includegraphics[scale=1.2]{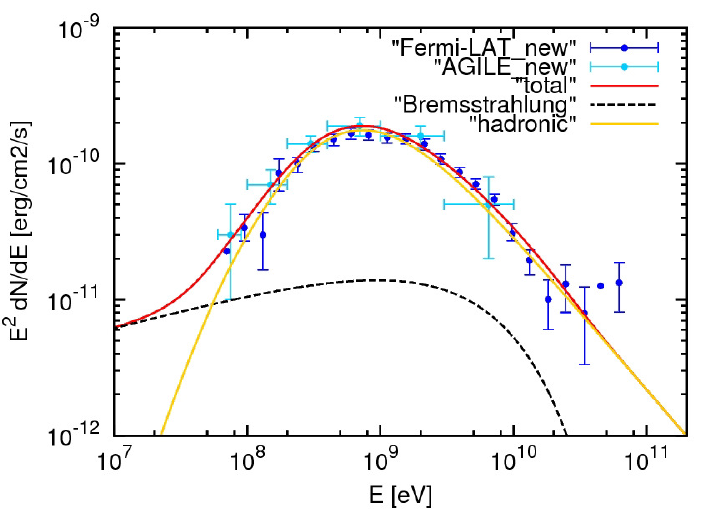}
\end{center}
\caption{Hadronic model H1 of the gamma-ray spectrum of 
W44 superimposed with the gamma-ray data of
Fig.~\ref{spectra} (in blue and cyan color). We find an index
$p_{1}=2.0\pm0.1$ with a high-energy cut-off at $E^{p}_{c}$= 45
GeV. This model is characterized by \textit{B} =210~$\mu$G and
\textit{n}=300~cm$^{-3}$. The yellow curve shows the neutral pion
emission from the accelerated proton distribution discussed in the
text. The black dashed curve show the electron
contribution by bremsstrahlung (dashed) emissions.  The IC
contribution is negligible. The red curve shows the total
gamma-ray emission from pion-decay and bremsstrahlung.}
\label{simpleHE}
\end{figure}

Following \cite{aharonian04}, we fit our W44 spectral data with a simple power-law with a high-energy cut-off.
In this case, our best-fitting parameters are an index
$p_{1}=2.0\pm0.1$,  and a cut-off energy at
$E^{p}_{c}=45$ GeV with $B=210$ $\mu$G. Deduced global
physical quantities are  a relatively large proton energy, $W_{p}=1.2\times10^{50}$ erg,
and a low electron/proton energy ratio $\Re=0.005$. Even in this
case, there are some points against the applicability of this model.
First, we can only fit our data with this proton distribution
ignoring the last four Fermi high energy points,  and
requiring a very low electron/proton ratio. Moreover, the
high-energy cut-off considered in \cite{aharonian04} for a SNR of similar age as W44 in
similar density and magnetic field conditions is less sharp than the one obtained in our
model, even with a high diffusion coefficient.

\subsection*{Smoothed broken power-law}

\begin{figure}[!h]
 \begin{center}
 \includegraphics[scale=1.2]{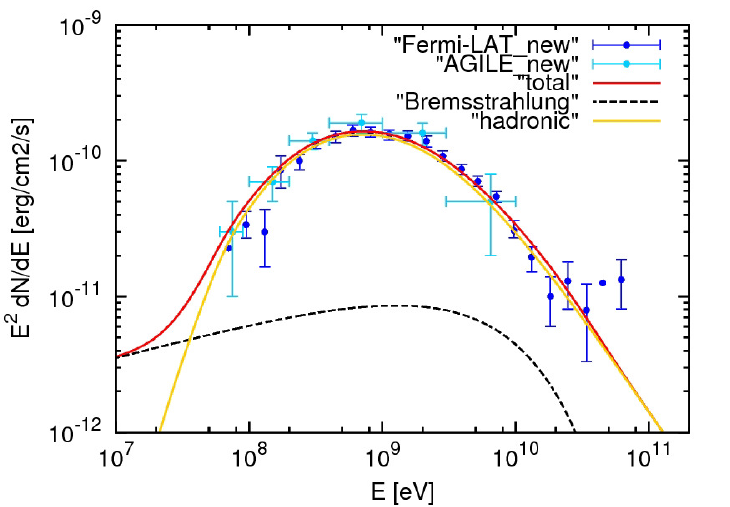}
\end{center}
\caption{Hadronic model H2 of the gamma-ray spectrum of 
W44 superimposed with the gamma-ray data of
Fig.~\ref{spectra} (in blue and cyan color). We find an index
$p_{1}=1.74.0\pm0.1$ (for $E<E_{br}$), $p_{2}=3.5\pm0.1$ (for
$E>E_{br}$) where \textit{$E^{p}_{br}$}= 16 GeV. This model is
characterized by \textit{B} =210~$\mu$G, and
\textit{n}=300~cm$^{-3}$. The yellow curve shows the neutral pion
emission from the accelerated proton distribution discussed in the
text. The black dashed curve shows the electron contribution by
bremsstrahlung (dashed) emissions; the IC contribution is
negligible. The red curve shows the total gamma-ray emission from
pion-decay and bremsstrahlung.} \label{brokenSM}
\end{figure}

Another way to  model the W44 gamma-ray spectral data is using a smoothed
broken power-law proton distribution (Eq.\ref{broken}). Our best
model provides indices 
$p_{1}=1.74\pm0.1$ for $E<E_{br}$, and
$p_{2}=3.5\pm0.1$ for $E>E_{br}$, with $E_{br}=16$ GeV, $B=210$
$\mu$G and $\Re=0.08$. We notice that for this proton
distribution, we obtain the same index of the electron
distribution. In our
opinion, however, the distribution of Eq.\ref{broken} cannot be considered a good
model. The reason is that this model introduces a strong
covariance between the low- and high-energy indices making their
determination quite difficult and questionable. We demonstrate
this point by showing in Fig. \ref{multiple_indices} the pion emission expected from proton
distribution of difference indices extending to the lowest
energies and with no breaks at higher energies.
The low-energy part of the spectrum of W44 can be
well fitted by an index in a range $2-2.3$. Steeper or harder indices cannot reproduce our data, 
this being true also for the value $p_{1}=1.74\pm0.1$ found with the approach considered here.

\subsection*{Simple power-law with a low-energy cut-off}

\begin{figure}[!h]
 \begin{center}
 \includegraphics[scale=1.2]{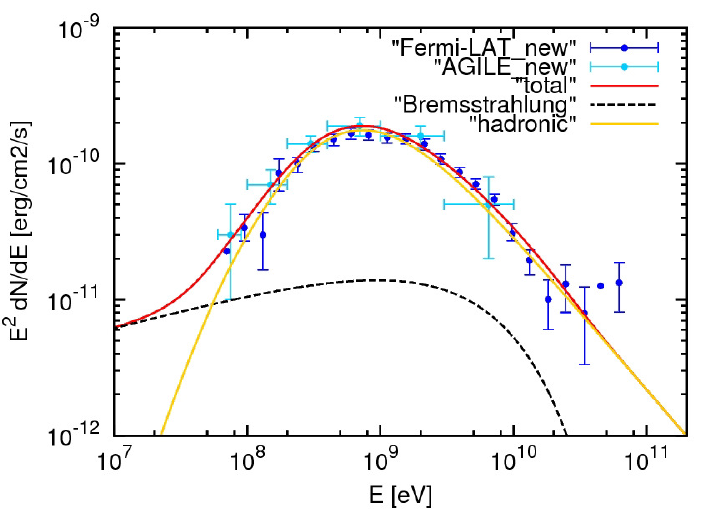}
\end{center}
\caption{Hadronic model H4 of the gamma-ray spectrum of 
W44 superimposed with the gamma-ray data of
Fig.~\ref{spectra} (in blue and cyan color). We find an index
$p_{1}=3.2\pm0.1$ with a low-energy cut-off at $E^{p}_{c}$= 6.5
GeV. This model is characterized by \textit{B} =145~$\mu$G and
\textit{n}=300~cm$^{-3}$. The yellow curve shows the neutral pion
emission from the accelerated proton distribution discussed in the
text. The black dashed curve show the electron contribution by
bremsstrahlung (dashed) emissions; the IC contribution is
negligible. The red curve shows the total gamma-ray emission from
pion-decay and bremsstrahlung.} 
\label{simple_LE}
\end{figure}

In our previous paper G11,
we used a simple power-law proton distribution in kinetic energy which resulted in a
spectral index $p_{1}=3.0\pm0.1$, and a low-energy cut off at
$E^{p}_{k,c}=5.5$ GeV.
We model here the W44 spectral data with the same kind of
distribution as in G11, but in total energy rather than kinetic
energy, as justified by the approach of  \cite{kelner06}:
\begin{equation}
 \frac{dN_{p,4}}{dE}= K_p E_{k}^{-p} e^{-\frac{E^p_{k,c}}{E}}
 \label{eq_simple_LE}
\end{equation}

We obtain a reasonable good modeling of the spectral data with an index $p_{1}=3.2\pm0.1$, and a cut-off energy
$E^{p}_{c}=6.5$ GeV, for a magnetic field $B=145$ $\mu$G, and an
electron/proton energy ratio $\Re=W_{e}/W_{p}=0.03$.
If the interaction of protons with the gas is outside the acceleration site,
the energy-depended diffusion of particles may produce this cutoff, as observed for the SNR W28 \citep{giuliani10_W28} and discussed by
\citep{gabici09}.
Alternatively, it may be due to a suppression of the  diffusion
coefficient due to turbulent motions in the cloud \citep{gabici07}
that would exclude the low-energy CRs from the denser regions.
In both cases a slow diffusion coefficient ($D\sim10^{26}$ cm$^{2}$s$^{-1}$  at 10 GeV)
is required.

\begin{figure}[!h]
 \begin{center}
 \includegraphics[scale=1.3]{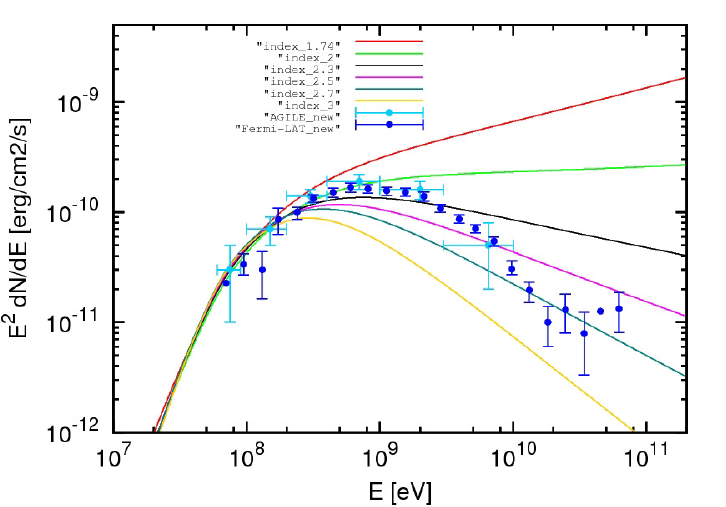}
\end{center}
\caption{Gamma-ray emission from neutral pion decay calculated for different
simple power-law proton distributions of different spectral indices without any break or
cut-off. We show the cases for
$p_{1}=1.74$ (red), $p_{1}=2$
(green), $p_{1}=2.3$ (black), $p_{1}=2.5$ (magenta), $p_{1}=2.7$
(cyan), and $p_{1}=3$ (yellow).}
\label{multiple_indices}
\end{figure}

\section*{Appendix C - Density and magnetic field link}
\label{appendixC}

\begin{figure}[ht!]
 \centering
 \subfigure{\includegraphics[scale=1.2]{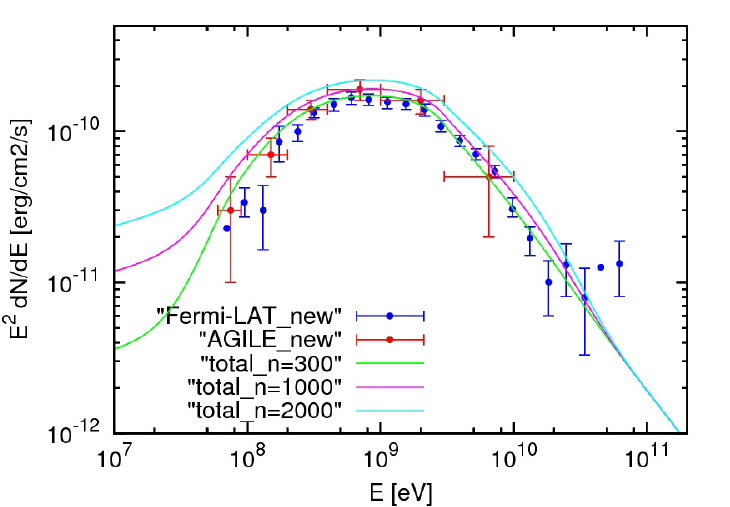}}
 \subfigure{\includegraphics[scale=1.2]{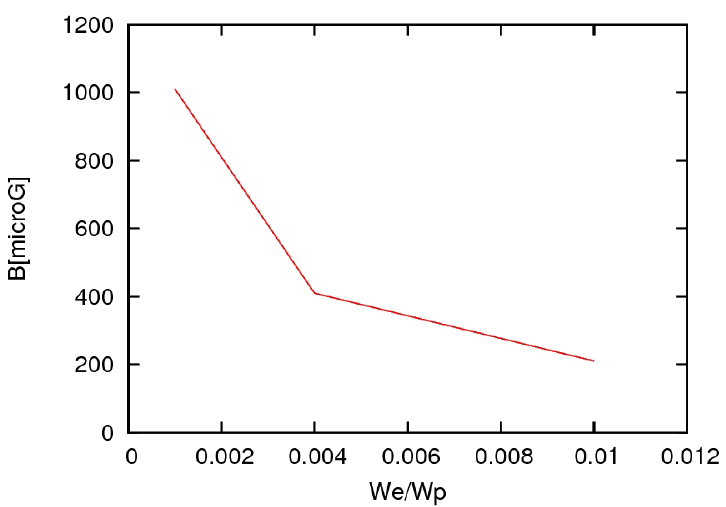}}
\caption{\textbf{(Top Panel)}: Our best hadronic model H3 
(see Fig.~\ref{hadronic}), for
three different values of the target density; n=300,
1000, 2000 cm$^{-3}$. \textbf{(Bottom Panel)}: correlation between the magnetic field  and the electron/proton ratio in order
to fit the spectral gamma-ray data of W44 for the
three assumed target densities. } 
\label{highdensity}
\end{figure}

The relation between the magnetic field and the target density in W44 is important for our modeling.
We consider here a hadronic model with the same parameters of our best model
(broken power-law distribution, see Table~\ref{table}) but with
higher density values, $n=1000$ cm$^{-3}$ and $n=2000$ cm$^{-3}$.
We assume that all gamma-ray emission detected by AGILE 
originates from the core of the W44 molecular cloud
at Galactic coordinates (34.75,-0.5). In
Fig.~\ref{highdensity} (top panel), we show the results obtained
with these high density values together with the one obtained for
$n=300$ cm$^{-3}$. The higher the density, the higher the bremsstrahlung contribution to the total gamma-ray
emission. A too large value of the target density may lead
to an overestimation of the high energy gamma-ray data. The only way to 
fit the data is by assuming a lower electron density
(and a lower electron/proton energy ratio $\Re=W_{e}/W_{p}$) with
a consequently higher magnetic field, $B\sim410$ $\mu$G and
$B\sim1110$ $\mu$G for $n=1000$ cm$^{-3}$ and $n=2000$ cm$^{-3}$,
respectively (Fig.~\ref{highdensity}, bottom panel).

\newpage

\begin{table*}[!h]
 \caption{Summary of the model parameters used in order to fit gamma-ray and radio W44 data. $<B>$ is the average magnetic field, $<n>$ is the average density, $E^{p}_{br}$ and $E^{e}_{br}$ are the proton and electron break energies, $E^{p}_{c}$ and $E^{e}_{c}$ are the proton and electron cut-off energies, $p_{1}$ and $p_{2}$ are the proton indices below and above the break, $p'_{1}$ and $p'_{2}$ are the electron indices below and above the break, $W_{p}$ and $W_{e}$ are the proton and electron total energies.}
\scriptsize
\centering
\label{table}
\begin{tabular}{l c c c c c c c c c c c c}
\hline\hline
\textbf{Models}              & \textbf{$<$B$>$} & \textbf{$<$n$>$}    & $\mathbf{E^{p}_{br}}$  & $\mathbf{E^{p}_{c}}$    & $\mathbf{E^{e}_{br}}$    &  $\mathbf{E^{e}_{c}}$   & \textbf{$p_{1}$} & \textbf{$p_{2}$} & \textbf{$p'_{1}$}   & \textbf{$p'_{2}$} & \textbf{$W_{p}$}    & \textbf{$W_{e}$}     \\
                             &$\mathbf{[\mu G]}$&$\mathbf{[cm^{-3}]}$ &   \textbf{[GeV]}       & \textbf{[GeV]}          & \textbf{[GeV]}           &  \textbf{[GeV]}         &                  &                  &                     &                         &  \textbf{[erg]}     &  \textbf{[erg]}         \\
\hline
                             &                  &                     &                        &                         &                          &                         &                  &                  &                     &                    &                     &                              \\
\textbf{\cite{GiuCaTa11}}    &         70       &      100            &           -            &   $5.5\pm1$ (LE)        &       -                  &       $15\pm1$          &    $3.0\pm0.1$   &     -            &    1.74               &           -           &$3.3\times10^{49}$   & $2.8\times10^{48}$    \\
                             &                  &                     &                        &                         &                          &                         &                  &                  &                     &                    &                     &           \\     
                             &                  &                     &                        &                         &                          &                         &                  &                  &                     &                    &                     &                        \\
\textbf{\cite{ackermannW44}} &         -        &   100               &      $22\pm1$          &        -                &       -                  &                         &    $2.36\pm0.05$ &  $3.5\pm0.3$     &   -              &             -                    &  $4\times10^{49}$   &          -             \\
                             &                  &                     &                        &                         &                          &                         &                  &                  &                     &                    &                     &                  \\  \cline{1-13}
\hline
                             &                  &                     &                        &                         &                          &                         &                  &                  &                     &                    &                             &             \\
 \textbf{H1}                 &     210          &  300                &       -                &  $45\pm1$ (HE)          &     -                    &          $20\pm1$       &    $2.0\pm0.1$   &   -              &     1.74            &  -                        & $1.2\times10^{50}$  &  $6.4\times10^{47}$     \\
                             &                  &                     &                        &                         &                          &                         &                  &                  &                     &                    &                     &                  \\  
                             &                  &                     &                        &                         &                          &                         &                  &                  &                     &                    &                     &                   \\
\textbf{H2}                  &     210          &   300               &       $16\pm1$         &  -                      &     -                    &          $15\pm1$       &   $1.7\pm0.1$    &   $3.5\pm0.1$    &      1.74           &  -            &  $1.3\times10^{49}$ &$9.6\times10^{48}$       \\
                             &                  &                     &                        &                         &                          &                         &                  &                  &                     &                    &                     &                   \\ 
                             &                  &                     &                        &                         &                          &                         &                  &                  &                     &                    &                     &                 \\
\textbf{H3}                  &  \textbf{210}    &   \textbf{300}      &    \textbf{20$\pm$1}   & \textbf{-}              &     -                    &      \textbf{12$\pm$1}  &\textbf{2.2$\pm$0.1}& \textbf{3.2$\pm$0.1} &\textbf{1.74}  &  \textbf{-}        & \textbf{$5\times10^{49}$}& \textbf{$5.6\times10^{47}$}       \\
                             &                  &                     &                        &                         &                          &                         &                  &                  &                     &                    &                     &                   \\
 \hline
                             &                  &                     &                        &                         &                          &                         &                  &                  &                     &                    &                     &                  \\
   \textbf{L1}               & 25               &     300             &                        &                         &         $8\pm1$          &           -             &       -          &    -             &    1.74             &     $4.2\pm0.1$    &                       -         &   $3.2\times10^{48}$                          \\
                             &                  &                     &                        &                         &                          &                         &                  &                  &                     &                    &                     &                           \\
                             &                  &                     &                        &                         &                          &                         &                  &                  &                     &                    &                     &                    \\
  \textbf{L2}                &    40            &      300            &         -              &  -                      &        $0.5\pm0.1$       &        -                &     -            &    -             &    $-2.5\pm0.1$     &  $3.4\pm0.1$         &          -          &     $6.6\times10^{47}$                 \\
                             &                  &                     &                        &                         &                          &                         &                  &                  &                     &                    &                     &                   \\

                             \hline
 \end{tabular}
\end{table*}

\end{document}